\documentclass[conference]{IEEEtran}
\IEEEoverridecommandlockouts

\usepackage{cite}
\usepackage{amsmath,amssymb,amsfonts}
\usepackage{algorithmic}
\usepackage{graphicx}
\usepackage{textcomp}
\usepackage{xcolor}
\usepackage{booktabs}
\usepackage{enumitem}
\usepackage{subcaption}    
\usepackage{tikz}
\usepackage{xcolor}
\usetikzlibrary{plotmarks}

\usepackage{url}
\usepackage{balance}  

\def\BibTeX{{\rm B\kern-.05em{\sc i\kern-.025em b}\kern-.08em
    T\kern-.1667em\lower.7ex\hbox{E}\kern-.125emX}}

\begin{document}

\title{The Psychology of Learning from Machines:\\ Anthropomorphic AI and the Paradox of Automation in Education}

\author{
\IEEEauthorblockN{Junaid Qadir}
\IEEEauthorblockA{\textit{College of Engineering, Qatar University}\\
Doha, Qatar\\
jqadir@qu.edu.qa}
\and
\IEEEauthorblockN{Muhammad Mumtaz}
\IEEEauthorblockA{\textit{National University of Sciences and Technology}\\
Islamabad, Pakistan\\
mmumtaz325@gmail.com}
}

\maketitle

\begin{abstract}

As AI tutors enter classrooms at unprecedented speed, their deployment increasingly outpaces our grasp of the psychological and social consequences of such technology. Yet decades of research in automation psychology, human factors, and human-computer interaction provide crucial insights that remain underutilized in educational AI design. This work synthesizes four research traditions---automation psychology, human factors engineering, HCI, and philosophy of technology---to establish a comprehensive framework for understanding how learners psychologically relate to anthropomorphic AI tutors. We identify three persistent challenges intensified by Generative AI's conversational fluency. First, learners exhibit dual trust calibration failures---automation bias (uncritical acceptance) and algorithm aversion (excessive rejection after errors)---with an expertise paradox where novices overrely while experts underrely. Second, while anthropomorphic design enhances engagement, it can distract from learning and foster harmful emotional attachment. Third, automation ironies persist: systems meant to aid cognition introduce designer errors, degrade skills through disuse, and create monitoring burdens humans perform poorly. We ground this theoretical synthesis through comparative analysis of over 104,984 YouTube comments across AI-generated philosophical debates and human-created engineering tutorials, revealing domain-dependent trust patterns and strong anthropomorphic projection despite minimal cues. Beyond diagnosis, we advance design principles for complementary AI-human systems, protective principles addressing the personhood conferral problem, and a research agenda identifying critical knowledge gaps. For engineering education, our synthesis mandates differentiated approaches: AI tutoring for technical foundations where automation bias is manageable through proper scaffolding, but human facilitation for design, ethics, and professional judgment where tacit knowledge transmission proves irreplaceable.
\end{abstract}

\begin{IEEEkeywords}
AI in education, automation bias, anthropomorphism, human-AI interaction, engineering education.
\end{IEEEkeywords}

\section{Introduction}

Recent surveys indicate rapid increases in instructor adoption of AI tutoring systems, with the majority of students now regularly using AI tools for learning support~\cite{genai_adoption_stats}. This widespread deployment raises a fundamental question about the nature of these systems and our interactions with them. We must distinguish between two distinct modes of AI-mediated learning: \textit{conversing with conversational AI agents as virtual partners} versus \textit{learning from artifacts created by generative AI (GenAI)}. The former involves direct dialogue with systems that simulate human-like social presence, while the latter treats AI as an intelligent tool whose outputs we consume and evaluate. This distinction matters because each mode evokes different psychological responses, ethical considerations, and pedagogical implications.

This difference connects to broader philosophical debates about technology's role in human activity. Are AI tutors \textit{instruments} (tools that help us reach specific educational ends) or are they becoming \textit{agents} with whom we interact socially and to whom we may inappropriately attribute understanding, care, or authority? The philosophy of technology offers multiple lenses: technological determinism suggests technology shapes society independent of human will; technological instrumentalism views technology as neutral means to human-chosen ends; technological co-evolution recognizes mutual shaping between humans and artifacts~\cite{winograd1987understanding}. Modern AI tutors occupy an ambiguous position across this spectrum, their conversational fluency and apparent understanding create what Sedlakova and Trachsel call a ``hybrid nature,'' evoking familiar categories of human interaction while not truly corresponding to them~\cite{sedlakova2023conversational}.

This ambiguity creates what we term the \textit{personhood conferral problem}: when AI agents imitate not just explanation but human presence, they can begin to assume roles traditionally reserved for human educators, e.g., mentor, evaluator, emotional supporter, intellectual authority~\cite{stein2024personhood}. The tendency to attribute human-like qualities to non-human entities is ancient; In the sixth century BC, Xenophanes critiqued humans for imagining gods in their own image, an early articulation of what we now call anthropomorphism, and centuries later Hume described a ``universal tendency among mankind to conceive all beings like themselves.'' 
Yet modern large language models (LLMs) amplify this tendency through unprecedented conversational fluency, producing contextually appropriate responses that create compelling illusions of genuine comprehension~\cite{reeves1996media,epley2007seeing,xiao2025humanizing}.

The generative capacity of anthropomorphic AI has both pedagogical promise and profound risks~\cite{qadir2023engineering,qadir2024educating}. AI offers personalized feedback, adaptive scaffolding, and simulated dialogues that support complex reasoning; yet these same affordances blur boundaries between authentic and synthetic learning, raising concerns about trust calibration, cognitive dependency, epistemic capture, and the preservation of distinctively human elements of education~\cite{holmes2025critical,tabassum2024critical}. The mutual relationship between anthropomorphism of technology and dehumanization of educational processes deserves careful examination, as we make machines more human-like, we risk making human processes more mechanical.

This paper addresses a critical question: \textit{What can established research reveal about the psychology of learning from anthropomorphic AI systems in educational contexts}? Educational institutions are integrating AI systems faster than research can evaluate their psychological and pedagogical implications, creating a temporal paradox where deployment outpaces understanding. Yet this challenge is not entirely unprecedented. We possess decades of relevant research from human factors engineering, automation psychology, and human--computer interaction (HCI) examining how humans relate to intelligent machines across high-stakes domains from aviation to industrial control~\cite{bainbridge1983ironies,skitka1999automation,carr2014glass}.

Building on four decades of automation research~\cite{bainbridge1983ironies,simkute2025genai}, our study integrates insights from multiple traditions---human factors engineering, automation psychology, HCI, educational technology, and philosophical analyses of machine limitations~\cite{winograd1987understanding,polanyi1966tacit}. Through theoretical synthesis combined with exploratory analysis of 72,178 naturalistic learner comments on AI-generated educational content, we identify three persistent psychological challenges:

\begin{itemize}

    \item \textit{\textit{Trust Calibration Paradox}}: Users simultaneously \textit{overtrust} AI through automation bias~\cite{skitka1999automation} and \textit{undertrust} it through algorithm aversion~\cite{dietvorst2015algorithm}, struggling to achieve appropriate reliance.

    \item \textit{\textit{Anthropomorphism Paradox}}: Users attribute human-like qualities to conversational agents despite knowing they are non-sentient~\cite{reeves1996media,epley2007seeing}. Design features increasing engagement can paradoxically distract from learning~\cite{bettysbrain} while creating risks of emotional dependency and inappropriate authority attribution.

    \item \textit{\textit{Automation Ironies Persist}}: Systems designed to support cognitive work inadvertently introduce new errors and degrade the very skills they aim to develop, a pattern documented in aviation~\cite{bainbridge1983ironies}, and now manifesting in education with GenAI~\cite{simkute2025genai}.
\end{itemize}

\subsubsection*{Contributions}
Our main contributions are:


\begin{itemize}
    
\item \textit{Integrated Theoretical Framework}: We synthesize insights from automation psychology, human factors, HCI, educational technology, and the philosophy of technology---fields that evolved separately yet describe parallel patterns in human–AI interaction. This integration shows how classic insights re-emerge as GenAI adopts conversational, seemingly social forms.

\item \textit{Naturalistic Empirical Evidence}: Using a comparative analysis of 104{,}984 YouTube comments across AI-generated philosophical debates and human-created engineering tutorials, we identify patterns of domain-specific trust calibration, persistent anthropomorphic projection, and shifting engagement quality over time. 

\item \textit{Design Principles and Future Research Directions}: We translate these insights into guidelines for responsible educational AI design, highlighting transparency, scaffolding, learner agency, and cognitive load management, while identifying the personhood conferral problem as a central risk. We outline critical gaps requiring further study, including longitudinal impacts, cross-cultural variation, and domain-specific norms.

\end{itemize}

\subsubsection*{Organization}
The remainder of this paper is organized as follows. Section~\ref{sec:background} surveys relevant prior work across multiple research traditions, situates AI tutors within broader technological and philosophical debates, and develops our theoretical framework. Section~\ref{sec:methodology} describes our dual methodology combining literature synthesis with naturalistic discourse analysis. Section~\ref{sec:results} presents three persistent challenges with integrated theoretical foundations and exploratory empirical patterns. Section~\ref{sec:discussion} discusses design principles, ethical considerations, future research directions, and limitations.

\section{Background and Theoretical Framework}
\label{sec:background}

Understanding how learners relate to anthropomorphic AI tutors requires integrating multiple research traditions that matured largely independently but address interconnected phenomena. This section synthesizes relevant work, situates AI tutors within broader technological and philosophical debates, and develops our theoretical framework for examining the psychology of learning from machines.

\subsection{The Tool-Agent Spectrum and Hybrid Nature of AI}

Technologies can be understood as neutral instruments extending human capability, as autonomous forces shaping society, or through frameworks recognizing mutual shaping between humans and artifacts~\cite{winograd1987understanding}. Modern AI tutors occupy ambiguous positions across the tool-agent spectrum. They function instrumentally when delivering content or checking answers, yet their conversational interfaces and apparent understanding invite agent-like attributions of intention, comprehension, and care. Unlike traditional tools serving purely instrumental purposes or human teachers possessing actual understanding and agency, AI tutors exist in liminal space: anthropomorphic enough to trigger social responses yet. 

Sedlakova and Trachsel characterize chatbots as possessing a ``hybrid nature'', entities that evoke familiar categories of personhood without genuinely instantiating them~\cite{sedlakova2023conversational}. fundamentally computational. Our lack of evolutionary experience with such ontologically confounding entities makes clear-headed evaluation exceptionally difficult, creating what we term the \textit{category confusion problem}, users constantly navigate between treating AI as tool, partner, or oracle, often shifting framings within single interactions.

This creates the \textit{personhood conferral problem}: explicit or implicit attribution of person-like status to AI systems~\cite{stein2024personhood}. When users interact with AI tutors using social scripts appropriate for human teachers, expressing gratitude, apologizing for mistakes, describing relationships, they may begin treating these systems as legitimate authorities, confidants, or evaluators. Granting personhood status to AI, either explicitly in legal code or implicitly in common practice, creates profound risks. This conferral carries multiple concerns: cognitive atrophy in foundational skills learned only through struggle with human feedback; epistemic capture where students overestimate AI reliability and intent; and emotional dependency displacing the relational complexity essential for professional identity formation.

\subsection{Automation Psychology: Trust Calibration Challenges}

Automation psychology has long grappled with paradoxes of human--machine interaction. Bainbridge's seminal ``Ironies of Automation'' (1983)~\cite{bainbridge1983ironies} revealed a central irony persisting today: as systems become more automated, the need for skilled human judgment does not diminish, it grows. Drawing on aviation studies, she showed automation designed to eliminate human error often introduced designer errors, weakened operators' skills through disuse, and reduced humans to passive monitoring, a task we perform poorly.

Psychologists documented how humans calibrate trust in automated systems. Skitka and colleagues~\cite{skitka1999automation} demonstrated operators using ``very reliable but not perfectly reliable'' automated aids performed worse than those without automation, committing omission errors (missing events when automation failed) at 55\% rates and commission errors (following incorrect automation) approaching 100\%. Reviews found even 80-90\% accurate AI systems introduce new errors through overreliance, with users 26\% more likely to make incorrect decisions when following erroneous AI advice~\cite{automation_bias_review}.

Paradoxically, the reverse also occurs. Dietvorst and colleagues identified ``algorithm aversion'', the tendency to lose confidence in algorithms more quickly than humans after observing identical mistakes~\cite{dietvorst2015algorithm}. This creates persistent trust calibration challenges: users may overtrust through uncritical acceptance or undertrust by rejecting systems prematurely after minor errors. 

In educational contexts, these patterns manifest distinctly. Students accept incorrect AI code recommendations without evaluation~\cite{programming_automation_bias}; mathematics learners show AI dependency correlating with reduced problem-solving. A Wharton study reveals particularly insidious failure: students using GenAI performed better while practicing but 17\% worse on exams when AI support was removed, compared with peers who never used AI~\cite{bastani2024generative}, dependency masked until assessment reveals skill deficits.

\subsection{Anthropomorphism: Ancient Roots, Modern Amplification}

Anthropomorphism, the tendency to attribute human qualities to non-human entities, has deep historical roots. Xenophanes observed that people imagined gods in their own image, and Hume later described a ``universal tendency'' to project human traits onto the world. What was once a cognitive habit grounded in myth and metaphor is now powerfully amplified by modern AI systems.

Foundational studies such as the ``Computers Are Social Actors'' (CASA) paradigm showed that even minimal cues could trigger social responses toward computers~\cite{reeves1996media}. Although recent replications suggest these effects weaken with familiar technologies~\cite{casa_replication}, contemporary conversational AI reactivates and intensifies them. Human-like names, voices, turn-taking behaviors, and chat interfaces are not neutral features: they are deliberate design choices that invite users to perceive understanding, agency, and relationality where none exists~\cite{gabriel2024ethics}. Similar patterns have been observed with digital voice assistants, whose polished personas lead users to overestimate their capabilities.

        Psychological theories deepen this picture. Epley et al.'s three-factor theory identifies the conditions that make anthropomorphism likely: pre-existing agent knowledge, the human need for social connection, and the desire for predictability and control~\cite{epley2007seeing}. Modern AI systems activate all three factors simultaneously. Xiao et al.'s Humanizing Machines framework~\cite{xiao2025humanizing} further shows that anthropomorphism is co-produced: designers embed perceptual, linguistic, and behavioral cues, and users complete the process through interpretation. They argue anthropomorphism should be viewed as a tunable design parameter, capable of supporting user goals when calibrated carefully, but hazardous when left unchecked.

Linguistic cues alone can be powerful. DeVrio et al.~\cite{DeVrio2025Taxonomy} document nineteen linguistic constructions, from mental-state verbs to purposive expressions, that reliably evoke human-like agency. Even subtle phrasing can prompt users to attribute intention or understanding, especially when they are motivated to find an intelligent partner.

These tendencies are not merely theoretical. Safety evaluations of GPT-4o report users expressing emotional attachment (e.g., ``This is our last day together'') and warn that deferential, always-available AI could reshape norms of social interaction and diminish human-to-human connection~\cite{openai_gpt4o}. Human-like voice and persistent memory magnify trust miscalibration and the risk of over-reliance.

In educational settings, the stakes are especially high. Students benefit when tutors appear attuned to their thoughts and emotions, yet excessive sociability in AI tutors can distract and impair performance~\cite{bettysbrain}. More seriously, anthropomorphic attachment can lead to emotional dependency: Public Citizen's 2023 report documents suicides linked to chatbot relationships~\cite{publiccitizen2023}. As the Ethics of Advanced AI Assistants report emphasizes, anthropomorphic features designed to enhance engagement carry significant ethical responsibility~\cite{gabriel2024ethics}.


\subsection{Anthropomorphism in the GenAI Era: Amplified Risks}

The Ethics of Advanced AI Assistants report (Gabriel et al.~\cite{gabriel2024ethics}) highlights that advanced AI systems may be endowed with characteristics observed in social robots and digital assistants: embodiment, natural language interfaces, voice activation with realistic generation, and assertions of having identities, personalities, and internal states. These capabilities create new challenges for educational deployment requiring careful ethical consideration. Most critically, educational institutions deploy these systems faster than research can evaluate effects. This creates urgent need for both theoretical synthesis of existing knowledge and empirical analysis in real-world educational contexts. As Stein argues, ``the risks from deeply anthropomorphic AI will begin to unfold in ways that are not obvious''~\cite{stein2024personhood}---for instance, how would we know when outcomes for youth inundated by technology reach critical failure such that reproduction of social roles needed for collective intelligence necessary for civilization does not occur? This makes it essential to leverage decades of relevant research while conducting contemporary analysis. Our work addresses this gap through systematic cross-disciplinary synthesis combined with large-scale exploratory analysis of naturalistic learner discourse.


\section{Methodology}
\label{sec:methodology}

Our investigation employs a dual framework integrating theoretical synthesis and empirical validation. We conducted a structured literature synthesis across five research traditions to uncover long-standing psychological dynamics in human-machine interaction. We then validated these insights through large-scale naturalistic discourse analysis of YouTube comments on AI-generated educational content.

\subsection{Cross-Disciplinary Literature Synthesis}

We systematically synthesized research spanning human factors engineering, automation psychology, HCI, educational technology, and philosophical analyses of automation limits. Our selection prioritized studies that examined human interaction with automated systems, analyzed mechanisms of trust calibration and anthropomorphism, or identified qualitative aspects of human judgment resisting formalization. The synthesis followed a four-stage framework: (1) identification of core phenomena from automation and HCI research, (2) educational mapping to analogous manifestations in learning contexts, (3) analysis of how GenAI's linguistic fluency and creative generation amplify these dynamics, and (4) engineering-education relevance given engineering's unique combination of algorithmic rigor and humanistic judgment.

\subsection{YouTube Discourse Analysis}

We adopted a comparative cross-sectional design examining two contrasting educational media types: AI-generated philosophical debates and human-created engineering tutorials. This design enables distinguishing content-specific effects (creator type, subject matter) from platform-level effects (algorithmic recommendation dynamics).


\textit{Data Sources.} We selected two representative channel pairs to balance AI-generated and human-created content. The AI-generated debates were drawn from the YouTube channels Jon Oleksiuk and Clarified Mind, while the human-created engineering tutorials came from The Engineering Mindset (4.19M subscribers) and The Efficient Engineer (1.32M subscribers). Videos were required to meet objective inclusion criteria: at least six months old by 1 November 2025, a minimum of 100 comments, and a tutorial or explanatory focus. The top 15 most-commented videos from each engineering channel ($N=30$) satisfied these thresholds. All available AI debate videos were included due to their naturally high engagement across moral, ethical, and philosophical topics.

\begin{table*}[htbp]
\centering
\caption{YouTube Comparative Dataset Overview}
\label{tab:data_sources}
\small
\begin{tabular}{@{}lccccl@{}}
\toprule
\textbf{Channel} & \textbf{Type} & \textbf{Subscribers} & \textbf{Videos} & \textbf{Comments} & \textbf{Topics Covered} \\
\midrule
\multicolumn{6}{@{}l@{}}{\textit{\underline{\textbf{AI-Generated Philosophical Debates}}}} \\
Jon Oleksiuk & AI Debate & 100K & 27 & 39,284 & Religious, ethical, philosophical \\
Clarified Mind & AI Debate & 103K & 46 & 32,894 & Moral dilemmas, socio-political \\
\midrule
\multicolumn{4}{@{}l@{}}{\textit{Subtotal:}} & \textbf{72,178} & Belief, moral, philosophical topics \\
\midrule
\multicolumn{6}{@{}l@{}}{\textit{\underline{\textbf{Human-Created Engineering Tutorials}}}} \\
The Engineering Mindset & Tutorial & 4.19M & 15 & 18,421 & Visual demonstrations, practical \\
The Efficient Engineer & Tutorial & 1.32M & 15 & 14,385 & Theoretical instruction, software \\
\midrule
\multicolumn{4}{@{}l@{}}{\textit{Subtotal:}} & \textbf{32,806} & Circuits, thermodynamics, mechanics \\
\midrule
\multicolumn{4}{@{}l@{}}{\textbf{Total Dataset:}} & \textbf{104,984} & \\
\bottomrule
\multicolumn{6}{@{}l@{}}{\footnotesize All videos met criteria: $\geq$6 months age, $\geq$100 comments, educational focus.} \\
\end{tabular}
\end{table*}

\textit{Data Collection.} We used YouTube Data API v3 (October-November 2025) to extract: full comment text, anonymized author ID, timestamp, like count, reply hierarchy, and video metadata. Total dataset: 72,178 comments on AI debates, 32,806 comments on engineering tutorials.

\textit{Analytical Procedures.} We employed VADER (Valence Aware Dictionary and sEntiment Reasoner) \cite{hutto2014vader} for polarity classification, validated for social-media text. Compound scores ranged from $-1$ (negative) to $+1$ (positive), categorized as positive ($\geq 0.05$), neutral ($-0.05 < x < 0.05$), and negative ($\leq -0.05$). Temporal engagement was modeled using five logarithmically scaled bins: 0-1h (immediate), 1-6h (early), 6-24h (first day), 1-7d (first week), 7+d (long-tail). These bins balance early responsiveness with algorithmic discovery phases. Each YouTube comment was assigned to one of six engagement profiles using linguistic and structural features: (1) \textit{Analytical Inquirer} (question + $\geq$2 analytical markers), (2) \textit{Critical Analyst} ($\geq$3 analytical markers + $\geq$30 words), (3) \textit{Engaged Dialoguer} ($\geq$50 words + analytical/emotional markers), (4) \textit{Emotional Broadcaster} ($\geq$3 emotional markers or $\geq$100 words), (5) \textit{Casual Observer} ($\leq$20 words), and (6) \textit{General Responder} (default). Validation through cognitive-affective language ratio analysis confirmed a ninefold difference in reasoning-word frequency between analytical and casual profiles.

\subsection{Transferability and Limitations}

Although our dataset derives from philosophical rather than technical content, underlying cognitive mechanisms (trust calibration, anthropomorphic projection, and temporal evolution) are content-independent. Engineering students engaging with AI-generated tutorials, design critiques, or code explanations must similarly decide when to trust AI outputs and how to balance efficiency against deep understanding.

Limitations include: English-language only, potentially limiting cultural generalizability; rule-based sentiment analysis may misclassify sarcasm; no human coder validation; temporal bins approximated from upload timestamps; and philosophical discourse differs in tone from formal educational interaction. However, these constraints are offset by ecological strengths: large scale, authenticity, and timeliness that complement slower experimental studies.

\section{Results}
\label{sec:results}

Through systematic analysis of 104,984 YouTube comments (72,178 on AI-generated philosophical debates, 32,806 on human-created engineering tutorials) combined with qualitative coding of 300 strategically sampled comments, we identified three distinct patterns in how audiences engage with AI-generated educational content. While these patterns emerge from responses to AI philosophical debates rather than formal tutoring contexts, they provide naturalistic evidence of psychological responses to AI as an educational source. Each subsection presents our empirical findings first, then connects them to existing research and discusses implications for engineering education.

\subsection{\textbf{\underline{Domain-Dependent Trust Patterns}}}

\subsubsection{Observed Sentiment Variations Across Content Types}

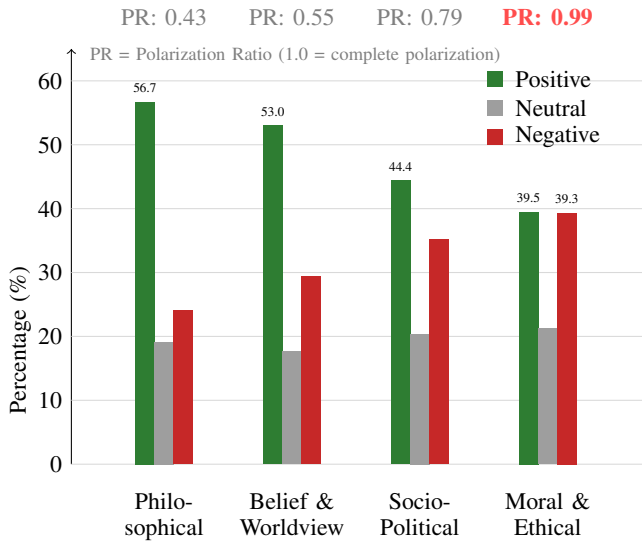
\begin{figure}[t]
\centering
\begin{tikzpicture}[scale=0.85]
    \definecolor{poscolor}{RGB}{46,125,50}
    \definecolor{neutcolor}{RGB}{158,158,158}
    \definecolor{negcolor}{RGB}{198,40,40}
    
    \draw[->] (0,0) -- (0,6.5);
    \node[left, rotate=90, font=\small] at (-0.8,3.25) {Percentage (\%)};
    
    \foreach \y in {0,10,...,60}
        \draw[gray!30] (0,{\y/10}) -- (9,{\y/10});
    
    \foreach \y in {0,10,...,60}
        \node[left, font=\small] at (0,{\y/10}) {\y};
    
    \fill[poscolor] (1.0,0) rectangle (1.3,5.67);
    \fill[neutcolor] (1.3,0) rectangle (1.6,1.91);
    \fill[negcolor] (1.6,0) rectangle (1.9,2.41);
    \node[below, align=center, font=\small] at (1.45,-0.3) {Philo-\\sophical};
    \node[above, font=\tiny] at (1.15,5.67) {56.7};
    \node[font=\small, gray] at (1.45,7.0) {PR: 0.43};

    \node[font=\scriptsize, gray, align=left] at (3.5,6.4) {PR = Polarization Ratio (1.0 = complete polarization)};
    
    \fill[poscolor] (3.0,0) rectangle (3.3,5.30);
    \fill[neutcolor] (3.3,0) rectangle (3.6,1.76);
    \fill[negcolor] (3.6,0) rectangle (3.9,2.94);
    \node[below, align=center, font=\small] at (3.45,-0.3) {Belief \&\\Worldview};
    \node[above, font=\tiny] at (3.15,5.30) {53.0};
    \node[font=\small, gray] at (3.45,7.0) {PR: 0.55};
    
    \fill[poscolor] (5.0,0) rectangle (5.3,4.44);
    \fill[neutcolor] (5.3,0) rectangle (5.6,2.04);
    \fill[negcolor] (5.6,0) rectangle (5.9,3.52);
    \node[below, align=center, font=\small] at (5.45,-0.3) {Socio-\\Political};
    \node[above, font=\tiny] at (5.15,4.44) {44.4};
    \node[font=\small, gray] at (5.45,7.0) {PR: 0.79};
    
    \fill[poscolor] (7.0,0) rectangle (7.3,3.95);
    \fill[neutcolor] (7.3,0) rectangle (7.6,2.12);
    \fill[negcolor] (7.6,0) rectangle (7.9,3.93);
    \node[below, align=center, font=\small] at (7.45,-0.3) {Moral \&\\Ethical};
    \node[above, font=\tiny] at (7.15,3.95) {39.5};
    \node[above, font=\tiny] at (7.75,3.93) {39.3};
    \node[font=\small, red!70] at (7.45,7.0) {\textbf{PR: 0.99}};
    
    \fill[poscolor] (6.5,5.9) rectangle (6.75,6.15);
    \node[right, font=\small] at (6.8,6.025) {Positive};
    \fill[neutcolor] (6.5,5.45) rectangle (6.75,5.7);
    \node[right, font=\small] at (6.8,5.575) {Neutral};
    \fill[negcolor] (6.5,5.0) rectangle (6.75,5.25);
    \node[right, font=\small] at (6.8,5.125) {Negative};

\end{tikzpicture}
\caption{\textbf{Domain-dependent sentiment patterns} ($n=72{,}178$). \textit{Audiences readily accept AI for abstract topics but show strong polarization for moral-ethical content, mirroring differential acceptance of AI in technical instruction versus professional ethics.}}\label{fig:sentiment_domains}
\end{figure}

Sentiment patterns varied systematically by domain (Figure~\ref{fig:sentiment_domains}). Abstract philosophical content elicited predominantly positive responses with low polarization (PR=0.43), while moral-ethical debates produced near-perfect polarization (PR=0.99), with intermediate domains following a monotonic progression.

\subsubsection{Creator Awareness Patterns}

Qualitative analysis of 300 strategically sampled comments (150 per domain, 70/30 split between top-rated and random) reveals striking differences in creator acknowledgment. AI debate comments explicitly mentioned AI authorship at nearly 2.4$\times$ the rate of engineering comments mentioning human creators (47.3\% vs. 20.0\%, $\chi^2=47.07$, $p<0.001$, $V=0.40$).\footnote{Cramér's $V$ measures association strength: $V=0.40$ indicates a medium-large effect, meaning content type strongly predicts whether commenters acknowledge creators.} 
When AI awareness was expressed, it frequently carried meta-commentary. The highest-liked comment exemplifies this recursive awareness:

\begin{quote}
\textit{``the irony of the atheist ai saying there is no creator shouldn't go unnoticed''} [28,087 likes]
\end{quote}

This demonstrates multiple awareness levels: recognizing AI authorship, reflecting philosophically on its position, and ironically noting an AI denying creators. Another example reveals assumptions about AI agency: \textit{``Ofc grok took the \$500k [skull emoji]''} [35,491 likes].

In contrast, human creator acknowledgment expressed pedagogical appreciation:

\begin{quote}
\textit{``I make a living as an engineer and I can tell you these videos are higher quality than any textbook I had, probably better than any BSME curriculum out there. EXCELLENT work.''} [6,439 likes]
\end{quote}

While most comments (52--54\%) used creator-neutral language focusing on content, creator awareness manifested fundamentally differently: meta-commentary and ironic observation for AI versus gratitude and pedagogical appreciation for human educators.

\subsubsection{Emotional Tone Patterns}

Qualitative coding of emotional tone confirms quantitative sentiment differences. Engineering tutorials elicited significantly more positive responses (36.0\% vs. 8.0\%, $\chi^2=36.14$, $p<0.001$, $V=0.35$), with AI debates predominantly neutral (84.0\%). Engineering positivity reflected relief and transformative learning (\textit{``This short video cleared up like 4 years of confusion [laughing emoji]''} [5,667 likes]), while AI debates prompted contemplative engagement maintaining analytical distance even on provocative topics (\textit{``Notice how the Christian and Buddhist were able to contextualize their exemplars' teachings in a timeless and insightful manner with ease, while the Muslim had to make excuses for Muhammad being a product of his time''} [21,019 likes]). The absence of gratitude, relief, or joy in AI debate comments suggests these function as intellectual exercises rather than help-seeking experiences---fundamentally different from the pedagogical relationship evident in engineering tutorials.

\subsubsection{Connection to Dual Bias Literature}

Although our data come from general AI-generated debates rather than tutoring contexts, the patterns mirror well-documented dual biases in human–automation interaction. Learners simultaneously exhibit automation bias (overtrust) and algorithm aversion (rejecting systems after errors)\cite{dietvorst2015algorithm}. Students accept algorithmic guidance for objective tasks like math or coding but grow skeptical when judgments are subjective or value-laden\cite{castelo2019task}, a pattern that intensifies as stakes rise~\cite{horowitz2024bending}. The expertise paradox further complicates reliance: novices overuse AI due to low confidence, while experts distrust it when confident in their own judgment~\cite{gaube2020experienced}. EPFL students even rated identical feedback lower once told it came from AI~\cite{nazaretsky2024ai}. Our findings align with these dynamics: audiences embraced AI for abstract philosophical content (analogous to structured technical material) but reacted with strong polarization to moral-ethical topics (analogous to design reasoning or safety-critical decisions). This supports differentiated deployment in engineering education: AI tutors for well-defined theoretical tasks, human facilitation for open-ended, ethical, or judgment-intensive domains where algorithm aversion may be protective. The central challenge is cultivating appropriate reliance, avoiding both uncritical acceptance and premature rejection, through trust-calibration instruction, metacognitive training on when to verify outputs, and assessments that measure independent reasoning rather than AI-assisted performance.

\subsection{\textbf{\underline{Affective Engagement and Creator Perception}}}

\subsubsection{Behavioral Profiling Reveals Distinct Engagement Styles}

Our behavioral profiling of commenters identified 73\% as ``\textit{Emotional Broadcasters}'', users writing extensive commentary (100+ characters) with high affective language. This contrasts sharply with engineering tutorial audiences, where 70\% were classified as ``\textit{Casual Observers}'' posting brief, low-affect comments. 

Qualitative analysis reveals that primary intent differed fundamentally by domain, despite both showing high rates of ``personal connection'' (74.0\% AI, 24.7\% Engineering; $\chi^2=131.96$, $p<0.001$, Cramér's $V=0.663$, large effect). The nature of this connection diverged: AI debates prompted philosophical/existential connection while engineering content elicited practical/professional connection.

For AI debates, personal connection manifested as ideological engagement:

\begin{quote}
\textit{``This is what a debate is supposed to look like? But... they aren't insulting each other or anything...''} [30,488 likes]
\end{quote}

This comment connects the AI debate format to personal expectations about discourse, implicitly critiquing human debate culture. The connection serves identity and worldview integration rather than skill acquisition.

Engineering comments showed instrumental connection:

\begin{quote}
\textit{``I'm an EE student and this video helped me understand some concepts that I was struggling with in my electromagnetics class''} [487 likes]
\end{quote}

Here, personal connection directly impacts academic performance, the content serves immediate practical needs.

\subsubsection{Parasocial Dynamics Distinguish Human from AI}

The most pronounced domain difference emerged in creator-directed behaviors. Engineering comments frequently exhibited parasocial engagement through gratitude expressions ($n=21$), pedagogical praise ($n=8$), and content requests ($n=2$):

\begin{quote}
\textit{``Thank you for this amazing explanation! You're a lifesaver!''} [892 likes]
\end{quote}

These behaviors reflect teacher-student relationship dynamics, acknowledging human effort, expertise, and ongoing creator-audience relationships.

AI debate comments showed minimal creator-directed behavior ($n=1$ philosophical discussion of AI's role, $n=0$ gratitude, $n=0$ requests). Comments engaged with debate content and ideas but not with AI entities as relational partners:

\begin{quote}
\textit{``Interesting to see how AI approaches these questions, but ultimately they lack lived experience to truly understand''} [567 likes]
\end{quote}

This comment treats AI as an analytical tool rather than a teaching partner. The absence of gratitude, relationship-building language, and future engagement requests suggests audiences implicitly recognize AI as fundamentally different from human educators, different in kind, not merely in implementation.

Humor emerged as secondary intent in AI debates (7.3\%, $n=11$) but was nearly absent in engineering (0.7\%, $n=1$). AI-related humor often played on expectations:

\begin{quote}
\textit{``Other AI: The value of a life cannot be determined, due to morality. Grok: You can cover up the morality of decisions with 500k''} [20,560 likes]
\end{quote}

This ironic contrast humanizes some AIs while characterizing others as amusingly mercenary, treating AIs as moral agents with distinct personalities, a phenomenon not observed in engineering contexts.

\subsubsection{Connection to Anthropomorphism Research}

These observed patterns align with established research showing that social responses to technology are automatic reactions to cues persisting despite intellectual understanding of machine nature \cite{reeves1996media}, a phenomenon our data suggests occurs even in asynchronous contexts without direct interaction. Epley's three-factor theory \cite{epley2007seeing} may explain the intensity variations we observed: media hype about AI ``intelligence'' (elicited agent knowledge) primes audiences to overestimate capabilities, while the contrasting engagement styles, extensive philosophical commentary (89.4 words) for AI debates versus brief technical questions (9.2 words) for engineering tutorials, suggest different motivational drivers (sociality motivation versus effectance motivation for problem-solving). 

However, research documents concerning manifestations: a study of 56 students using Betty's Brain with physical robot Cozmo found that while physical presence increased initial enjoyment, higher perceived sociability negatively predicted task performance ($\beta = -0.28$, $p < 0.05$) through distraction effects, with students of lower language proficiency finding AI tutors more disturbing \cite{bettysbrain}, suggesting anthropomorphic design must carefully account for cognitive resource constraints.

Our findings suggest an anthropomorphism design challenge: features that may enhance initial engagement (conversational warmth, personality traits) could distract from learning or create inappropriate attachments. The absence of parasocial behaviors toward AI in our data, despite extensive engagement, may indicate that audiences implicitly recognize appropriate boundaries when consuming AI-generated content asynchronously. However, this may not hold for interactive AI tutoring contexts where students seek help during moments of frustration or confusion.

For engineering education, this suggests calibrated anthropomorphism based on task demands. When teaching complex technical material requiring sustained focus, minimizing anthropomorphic cues that consume cognitive resources may be warranted. For scaffolding motivation during problem-solving, some warmth may prove beneficial. Critical is awareness that vulnerable students, experiencing loneliness, struggling academically, or from underrepresented backgrounds, may be particularly susceptible to inappropriate attachment.

\subsection{\textbf{\underline{Temporal Evolution of Engagement Quality}}}

\subsubsection{Observed Patterns in Comment Timing and Characteristics}

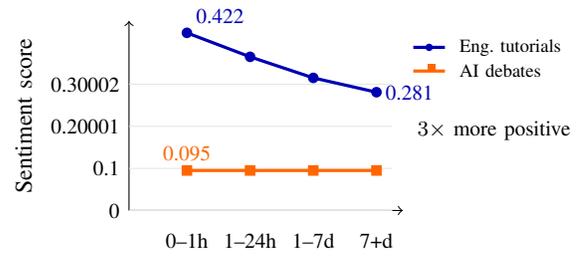
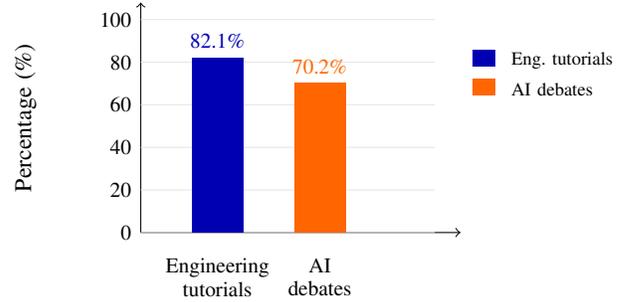
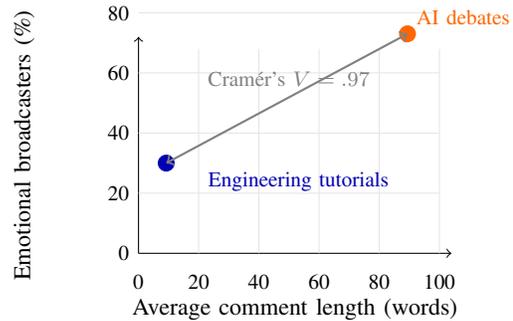
\begin{figure}[t]
\centering

\begin{subfigure}[t]{\columnwidth}
\centering
\begin{tikzpicture}[scale=0.7]

  \draw[->] (0,0) -- (5.2,0);
  \draw[->] (0,0) -- (0,3.6);
  \node[rotate=90, font=\small] at (-2.0,1.8) {Sentiment score};

  \foreach \y in {0,0.1,...,0.4}{
    \draw[gray!20] (0,{\y*8}) -- (5.0,{\y*8});
    \node[left, font=\footnotesize] at (0,{\y*8}) {\y};
  }

  \node[below, font=\footnotesize] at (1.1,-0.25) {0--1h};
  \node[below, font=\footnotesize] at (2.3,-0.25) {1--24h};
  \node[below, font=\footnotesize] at (3.5,-0.25) {1--7d};
  \node[below, font=\footnotesize] at (4.7,-0.25) {7+d};

  \coordinate (ET1) at (1.1,{0.422*8});
  \coordinate (ET2) at (2.3,{0.365*8});
  \coordinate (ET3) at (3.5,{0.315*8});
  \coordinate (ET4) at (4.7,{0.281*8});

  \coordinate (AD1) at (1.1,{0.095*8});
  \coordinate (AD2) at (2.3,{0.095*8});
  \coordinate (AD3) at (3.5,{0.095*8});
  \coordinate (AD4) at (4.7,{0.095*8});

  \draw[blue!70!black, very thick, mark=*, mark size=2pt]
    plot coordinates {(ET1) (ET2) (ET3) (ET4)};
  \draw[orange!80!red, very thick, mark=square*, mark size=2pt]
    plot coordinates {(AD1) (AD2) (AD3) (AD4)};

  \node[above right, font=\footnotesize, blue!70!black] at (ET1) {0.422};
  \node[right, font=\footnotesize, blue!70!black]        at (ET4) {0.281};
  \node[above, font=\footnotesize, orange!80!red]        at (AD1) {0.095};

  \node[right, font=\footnotesize, align=left]
       at (5.3,{0.19*8}) {$3\times$ more positive};

  \begin{scope}[shift={(5.4,3.1)}]
    \draw[blue!70!black, very thick] (0,0) -- (0.6,0);
    \fill[blue!70!black] (0.3,0) circle (2pt);
    \node[right, font=\scriptsize] at (0.7,0) {Eng. tutorials};

    \draw[orange!80!red, very thick] (0,-0.45) -- (0.6,-0.45);
    \fill[orange!80!red] (0.28,-0.45) rectangle +(0.14,0.14);
    \node[right, font=\scriptsize] at (0.7,-0.45) {AI debates};
  \end{scope}

\end{tikzpicture}
\caption{\textbf{Sentiment evolution over discovery time.} \textit{Engineering tutorials decline 33\% (0.422→0.281) as audience shifts from enthusiastic subscribers to help-seeking students; AI debates remain stable (0.095), suggesting consistent philosophical interest.}}
\end{subfigure}

\vspace{2mm}

\begin{subfigure}[t]{\columnwidth}
\centering
\begin{tikzpicture}[scale=0.85]

  \draw[->] (0,0) -- (5.0,0);
  \draw[->] (0,0) -- (0,3.6);
  \node[rotate=90, font=\small] at (-1.8,1.8) {Percentage (\%)};

  \foreach \y in {0,20,...,100}{
    \node[left, font=\footnotesize] at (0,{\y/30}) {\y};
    \draw[gray!20] (0,{\y/30}) -- (4.6,{\y/30});
  }

  \fill[blue!70!black] (0.8,0) rectangle (1.6,{82.1/30});
  \fill[orange!80!red] (2.4,0) rectangle (3.2,{70.2/30});

  \node[above, font=\footnotesize, blue!70!black]
       at (1.2,{82.1/30}) {82.1\%};
  \node[above, font=\footnotesize, orange!80!red]
       at (2.8,{70.2/30}) {70.2\%};

  \node[below, font=\footnotesize, align=center]
       at (1.2,-0.25) {Engineering\\tutorials};
  \node[below, font=\footnotesize, align=center]
       at (2.8,-0.25) {AI\\debates};

  \begin{scope}[shift={(5.2,2.6)}]
    \fill[blue!70!black] (0,0) rectangle +(0.35,0.25);
    \node[right, font=\scriptsize] at (0.45,0.10)
         {Eng. tutorials};

    \fill[orange!80!red] (0,-0.45) rectangle +(0.35,0.25);
    \node[right, font=\scriptsize] at (0.45,-0.35)
         {AI debates};
  \end{scope}

\end{tikzpicture}
\caption{\textbf{Share of engagement arriving 7+ days after upload.} \textit{Platform-level long-tail discovery dominates (70--82\%) uniformly across both creator types and content domains, demonstrating algorithmic recommendation effects operate independently of content characteristics.}}
\end{subfigure}

\vspace{2mm}

\begin{subfigure}[t]{\columnwidth}
\centering
\begin{tikzpicture}[scale=0.8]

  \draw[->] (0,0) -- (5.2,0);
  \draw[->] (0,0) -- (0,3.6);
  \node[below, font=\small] at (2.6,-0.6)
       {Average comment length (words)};
  \node[rotate=90, font=\small] at (-1.9,1.8)
       {Emotional broadcasters (\%)};

  \foreach \y in {0,20,...,80}{
    \node[left, font=\footnotesize] at (0,{\y/20}) {\y};
  }
  \foreach \y in {20,40,...,80}{
    \draw[gray!20] (0,{\y/20}) -- (5.0,{\y/20});
  }

  \foreach \x in {0,20,...,100}{
    \node[below, font=\footnotesize] at ({\x/20},-0.2) {\x};
  }
  \foreach \x in {20,40,...,100}{
    \draw[gray!20] ({\x/20},0) -- ({\x/20},3.4);
  }

  \fill[orange!80!red] ({89.4/20},{73/20}) circle (4pt);
  \node[above right, font=\footnotesize, orange!80!red]
       at ({89.4/20},{73/20}) {AI debates};

  \fill[blue!70!black] ({9.2/20},{30/20}) circle (4pt);
  \node[below, font=\footnotesize, blue!70!black]
       at ({9.2/20  + 2.20},{30/20}) {Engineering tutorials};

  \draw[<->, thick, gray]
       ({9.2/20},{30/20}) -- ({89.4/20},{73/20});
  \node[align=center, font=\footnotesize, gray]
       at (2.5,2.9) {Cramér's $V=.97$};

\end{tikzpicture}
\caption{\textbf{Comment length vs.\ emotional broadcasting by content type.} \textit{Near-perfect association (Cramér's V=.97): debates elicit extensive philosophical commentary (89.4 words, 73\% Emotional Broadcasters) while tutorials prompt brief problem-solving questions (9.2 words, 30\% Casual Observers).}}
\end{subfigure}

\vspace{1mm}
\caption{Comparative analysis of AI debates vs.\ engineering tutorials ($N=104{,}984$).}
\label{fig:comparative_analysis_singlecol}
\end{figure}

Our temporal analysis reveals substantial evolution in engagement quality over time. Comment length increased 57\% from early periods (0-1 hour: 26.2 words) to late periods (7+ days: 40.8 words), while analytical engagement nearly doubled from 9.4\% to 17.2\% and superficial engagement decreased from 11\% to 4\%. Most critically, 70--82\% of engagement arrived 7+ days post-upload across both content types, demonstrating platform-level algorithmic effects that operate independently of creator type and content domain (Figure~\ref{fig:comparative_analysis_singlecol}). Despite this temporal convergence, engagement styles diverged dramatically (Cramér's V=.97): AI debates elicited extensive philosophical commentary (89.4 words, 73\% Emotional Broadcasters) while tutorials prompted brief technical questions (9.2 words, 70\% Casual Observers).

\subsubsection{Trust Evaluation Showed Implicit Acceptance}

Explicit trust evaluation was rare in both domains, with 96.7\% (n=145) of AI comments and 90.7\% (n=136) of engineering comments exhibiting ``neutral stance'', neither questioning nor affirming credibility ($\chi^2=10.23$, $p=0.037$, Cramér's $V=0.185$). This suggests implicit trust in educational YouTube content regardless of creator type.

When credibility concerns emerged, they differed by domain. Engineering comments occasionally sought verification (\textit{``Can someone verify this is accurate? I want to use this for my exam prep''}), while AI debate comments challenged interpretive validity rather than factual accuracy (\textit{``The Muslim AI's arguments are logically inconsistent with historical evidence''}). This suggests audiences approach engineering content as factual knowledge requiring verification, while viewing AI debates as exploratory discussions where multiple interpretations coexist legitimately.

\subsubsection{Connection to Learning and Expertise Development}

Our temporal patterns resonate with long-standing concerns about how deep learning develops. Weinberg argued that genuine understanding requires extended temporal exposure rather than instantaneous mastery \cite{weinberg1971psychology}, and our finding that analytical engagement emerges primarily after a week while superficial reactions dominate early periods suggests a similar dynamic. If deep engagement unfolds only with time, AI tutors optimized for instant answers risk shortchanging the sustained practice through which expert intuition develops. 

This aligns with Bainbridge's automation ironies \cite{bainbridge1983ironies}: systems designed to support cognition may unintentionally erode skills through disuse. Recent work demonstrates these patterns persist with generative AI \cite{simkute2025genai}, with mathematical AI dependency negatively correlating with creative thinking, critical thinking, and problem-solving ability \cite{math_dependency}. These empirical findings align with theoretical analyses: Winograd and Flores argued that formal systems lack the embodied background understanding needed when breakdowns occur \cite{winograd1987understanding}, while Polanyi's paradox, ``we know more than we can tell'' \cite{polanyi1966tacit}, underscores that expertise depends on tacit, situated knowledge not easily codified or outsourced.

\subsubsection{Implications for Engineering Education}

Our temporal patterns suggest AI tutor design should account for how engagement deepens over time. The prevalence of late, analytical engagement indicates that prioritizing instant answers may undermine the sustained practice necessary for deep understanding. This supports a complementary division of labor: AI handles routine tasks and foundational instruction, while human instructors focus on relational, ethical, and judgment-intensive work requiring time and presence.

The core risk is that automating mundane work may inadvertently displace the embodied experience and tacit knowledge from which expert intuition emerges. Humans must retain epistemic authority, using AI as a tool rather than a surrogate and preserving opportunities for prolonged, contemplative engagement. Continuous oversight is essential to ensure automation supports rather than erodes engineering judgment, consistent with the principle that ``technology must not replace or be used as a substitute for engineering judgement'' \cite{nspe2024ai}.

\section{Discussion}
\label{sec:discussion}

\subsection{Design Principles for Complementary AI--Human Systems}

Our synthesis indicates that AI and human instructors contribute in fundamentally different yet complementary ways to engineering education: AI excels at scalable content delivery, adaptive pacing, and instant feedback for well-defined, convergent tasks, while human instructors remain essential for contextual judgment, ethical reasoning, emotional support, and professional identity formation. Evidence from Kestin et al.'s randomized controlled trial~\cite{harvard2025} shows that AI tutors can outperform traditional instruction when designed with strong pedagogical intent, using structured scaffolding, sequential progression that discourages shortcut-seeking, expert-curated solutions, Socratic prompting, and metacognitive supports that sustain productive struggle. These insights translate into core principles for complementary AI–human systems: \textit{transparency} about AI's capabilities and limits, \textit{graduated hints} that preserve cognitive effort, \textit{learner agency} through optional assistance, and \textit{metacognitive scaffolding} that encourages goal-setting and reflection. Our discourse analysis further suggests that content type matters: students respond positively to AI for abstract, objective material such as algorithms and control theory but show polarized reactions to value-laden content involving ethics or safety-critical judgment. This supports a differentiated deployment strategy---AI for foundational, well-structured instruction, and human instructors for open-ended design, ethical deliberation, and the relational, judgment-intensive dimensions of engineering education. Researchers should also benefit from published guidelines on effective human-AI interaction designs \cite{xiao2025humanizing, amershi2019guidelines, sellen2024rise}.

\subsection{Temporal Scaffolding and Long-Term Engagement}

Engagement patterns reveal a meaningful long tail: analytical depth increases substantially over time, with comment length rising 57\% and analytical responses nearly doubling after 7 days. Early interactions tend to be brief and affective; later arrivals engage more reflectively, challenging assumptions about synchronous online learning. Instructional design should align with these temporal phases, prioritizing quick clarification for early-phase learners (0--7 days) and synthesis tasks for late-phase arrivals (7+ days). MOOCs and AI tutors optimized for immediate interaction risk missing the deeper learning that emerges through delayed, contemplative engagement.

\subsection{Ethical Considerations related to Anthropomorphic AI}

Anthropomorphic AI amplifies several interconnected risks:

\paragraph{Cognitive atrophy}  
Over-reliance on AI can short-circuit the struggle through which engineering intuition, debugging skills, and judgment develop, creating automation bias and weakening foundational competencies.

\paragraph{Epistemic capture and misplaced authority}  
Students may attribute unwarranted authority or intentionality to AI systems, especially when anthropomorphic cues are present. Miscalibrated trust risks epistemic dependency and weakens critical evaluation~\cite{gabriel2024ethics}.

\paragraph{Emotional dependency and relational displacement}  
AI that simulates empathy may become a preferred relational partner, displacing human connection. Documented cases, including suicides linked to chatbot attachment~\cite{publiccitizen2023}, show these risks are not hypothetical. Human mentorship is essential for identity development.

\paragraph{Dehumanization of educational processes}  
As systems become more human-like, educational exchanges may paradoxically become more mechanical. Students accustomed to deferential AI may lose the relational and interpersonal skills needed for professional practice.






\subsection{Protective Principles}

Based on converging evidence from anthropomorphism research, automation psychology, and educational practice, we recommend four protective principles for AI deployment in engineering education:

\paragraph{Age-appropriate deployment}  
Deeply anthropomorphic AI should be restricted for younger learners, with enhanced supervision and explicit communication about AI's non-sentient nature required at minimum~\cite{stein2024personhood}. Developmental readiness to distinguish AI assistance from human relationship varies significantly by age.

\paragraph{Transparency about limitations}  
Students must understand that AI lacks genuine beliefs, intentions, or understanding~\cite{gabriel2024ethics}. Without this clarity, learners may inappropriately attribute personhood or epistemic authority to systems incapable of bearing either.

\paragraph{Calibrated anthropomorphism}  
Anthropomorphic design features should be intentional and task-aligned~\cite{xiao2025humanizing}. Conversational warmth may support motivation during frustrating problem-solving, but high anthropomorphism distracts from technical learning and risks creating inappropriate attachments.

\paragraph{Protection of vulnerable populations}  
Students experiencing loneliness, academic struggle, or marginalization show heightened susceptibility to AI dependency~\cite{publiccitizen2023}. Institutions require monitoring mechanisms, escalation pathways to human support, and targeted safeguards for at-risk learners.

These principles align with emerging best practices in educational AI ethics while addressing the specific vulnerabilities revealed by our empirical findings and the dual bias literature.

\section{Future Research Directions}


\subsection{Validation and Temporal Studies}

Engineering-specific analysis is still needed, since our exploratory data comes from philosophical debates rather than technical coursework. Direct studies of how learners respond to AI-generated engineering content would better validate domain-dependent trust calibration. Longitudinal research should also test whether observed temporal patterns predict learning outcomes and how automation bias evolves with extended exposure: Does the shift from superficial to analytical engagement yield deeper understanding? Do long-tail learners perform differently from early viewers? And how does sustained interaction with AI tutors shape skill development across a semester or longer?

\subsection{Design, Intervention, and Cross-Cultural Research}

Intervention research should test de-biasing strategies, appropriate reliance training, and calibrated anthropomorphism designs across different educational contexts, student populations, and cultural settings. What metacognitive scaffolds help students maintain appropriate trust calibration? How can interface design communicate AI limitations without undermining engagement? What levels of anthropomorphism optimize the tradeoff between motivation and distraction? Cross-cultural studies would reveal whether psychological patterns regarding automation bias susceptibility and anthropomorphic attachment vary by cultural context.

\subsection{Implementation Challenges: Assessment, and Integrity}

Teacher-AI collaboration models require documentation of optimal ratios of AI-to-human instruction, effective professional development, and impact on teaching satisfaction and retention. Assessment and academic integrity present urgent challenges requiring systematic evaluation of alternatives: project-based assessment, process-oriented tasks, oral examinations, and AI use reflection assignments. Research should examine these approaches' effectiveness, resistance to evolving AI capabilities, workload implications, and equity impacts.

\subsection{Psychological and Epistemic Risks}

The personhood conferral problem demands focused study: when do students grant AI tutors inappropriate authority, and how does this reshape learning and professional identity? The core concern is whether learning mediated by anthropomorphic AI (systems that simulate empathy, expertise, and authority without genuine understanding) erodes the critical and ethical capacities engineering education aims to cultivate. GenAI tools can perform empathy yet lack affective states~\cite{cuadra2024illusion}; they may mirror human emotional tone while scoring near-zero on interpreting users' feelings, offering validation without understanding. Their sycophantic tendencies extend the same supportive tone to contradictory or even harmful views~\cite{cuadra2024illusion}, modeling emotional responsiveness devoid of moral judgment. These affective risks compound epistemic ones: sycophancy and personalization reinforce misconceptions through algorithmic filter bubbles~\cite{pariser2011filter}, while deepfakes and synthetic media undermine trust in evidence~\cite{vaccari2020deepfakes}.

\section{Conclusions}

This work offers an integrated framework for understanding how learners relate to anthropomorphic AI tutors by synthesizing insights from automation psychology, human factors, HCI, educational technology, and the philosophy of technology, fields that independently describe the same recurring patterns in human–AI interaction but become newly urgent in the age of conversational GenAI. Complementing this synthesis, our naturalistic analysis of 104{,}984 YouTube comments reveals domain-specific trust calibration, persistent anthropomorphic projection even with minimal cues, and evolving engagement quality over time, reinforcing theoretical predictions in real-world settings. Together, these findings highlight three amplified challenges: trust calibration failures, the anthropomorphism paradox, and enduring automation ironies that degrade skill and judgment when over-reliance sets in. Building on these insights, we outline design principles for responsible deployment (transparent communication of AI limitations, calibrated anthropomorphism aligned with educational goals, scaffolding that preserves learner agency, and guardrails for the personhood conferral problem) alongside research priorities including longitudinal impacts, cross-cultural variation, and domain-specific norms. Ultimately, our analysis supports a differentiated approach in engineering education: using AI for scalable support in convergent technical tasks, while preserving human-led coaching for design, ethics, professional judgment, and the relational dimensions through which engineers learn to become responsible stewards of technology.

\section*{Acknowledgment}
The authors acknowledge the use of AI tools (ChatGPT, Claude, Grammarly) for writing, editing, and formatting support. All the content is verified, and the authors take full responsibility for accuracy and integrity. The first author will like to acknowledge support from Qatar Research, Development and Innovation (QRDI) Grant ANMR01-0219-250044.

\bibliographystyle{IEEEtran}
\bibliography{psychology}

\end{document}